\documentstyle[multicol,prl,aps,epsf]{revtex}
\begin{document}
\title{Intermolecular effects in the center-of-mass 
dynamics of unentangled polymer fluids}
\author{M. Guenza}
\address{
Department of Chemistry and Institute of Theoretical Science, 
University of Oregon, Eugene, OR 97403}
\maketitle
\begin{abstract}
The Generalized Langevin Equation for the Cooperative
Dynamics
of interacting polymer chains (M.Guenza, ${\it J.Chem.Phys.}$
v.$110$, $7574$ ($1999$))
is implemented to investigate the
anomalous dynamics of unentangled polymer melts. 
The proposed equation of motion formally relates the anomalous center-of-mass
diffusion, as observed in computer simulations and experiments, 
to the nature of the effective intermolecular mean-force potential.
An analytical Gaussian-core form of the potential between the centers
of mass
of two polymers is derived, which agrees with computer simulations
and allows the analytical solution of the equation of motion.
The calculated center-of-mass dynamics is characterized
by an initial subdiffusive
regime that persists for the spatial range of the intermolecular
mean-force potential, and for time intervals shorter than the first intramolecular
relaxation time, 
in agreement with 
experiments and computer simulations of unentangled polymer melt dynamics.
\end{abstract} 

\section{Introduction}
The dynamics of simple fluids (e.g. monoatomic or colloidal) is driven by the 
fluid-mediated interactions between the 
particles as they move under the influence of the "effective"
potential of mean-force.
The potential of mean-force is related to
the "static" radial distribution function $g(r)$, which describes
the probability of finding another particle at some distance $r$ from 
the first "tagged" particle. The structure of the fluid, and the shape of $g(r)$,
depends on the strength and shape of the "bare" interparticle potential, so
that the nature of the interparticle interactions completely
defines the static and dynamical properties of the fluid.\cite{hansen}

For molecular fluids, the liquid structure and dynamics
become more complex 
as they are driven by 
the interplay between intermolecular and intramolecular
interactions.\cite{hansen,rism} However,
for high molecular weight fluids such as polymer melts, the description is
again conventionally simplified by assuming that 
the intramolecular forces dominate 
the intermolecular forces due to chain connectivity.
In this case, the dynamics of a single chain is described by
a 
Generalized Langevin Equation (GLE) where the 
surrounding fluid acts as a  mean-field medium that interacts with the 
"tagged" molecule only through
the effective friction coefficient and the random intermolecular forces. Specifically,
unentangled
polymer fluids are described by the Rouse equation, 
an "intramolecular"  GLE where the 
memory function is neglected.\cite{BixonZw}

The Rouse equation has proved very 
successful in explaining several experimental findings 
of unentangled polymer fluid dynamics.\cite{DoiEdw}
It correctly predicts the scaling 
with degree of polymerization of the diffusion coefficient and bulk 
viscosity. 
Nevertheless, it appears to fail in describing
the center-of-mass (c.o.m.) anomalous dynamics experimentally observed for 
$t \leq \tau_{Rouse}$, with $\tau_{Rouse}=R_g^2/D_{Rouse}$ 
the first Rouse relaxation time. Here $R_g=l\sqrt{N/6}$ is
the molecular radius-of-gyrations, with $N$ the degree of 
polymerization, and $l$ the statistical segment length.
The Rouse diffusion coefficient $D_{Rouse}=k_B T/(N \zeta )$, where
$k_B$ is Boltzmann's constant, $T$ is the absolute temperature,
and $\zeta$ is the Rouse
monomer friction coefficient.
In the short-time regime ($t\le \tau_{Rouse}$) the Rouse equation predicts linear-in-time
purely diffusive c.o.m. dynamics, while
the observed c.o.m. mean-square displacement 
follows a
subdiffusive regime  ($\Delta R^2(t) \propto t^{\nu}$), with $\nu \approx 0.75 - 
0.9$, which
crosses-over to the free Rouse diffusion
($\Delta R^2(t) \propto t$) at $t \approx \tau_{Rouse}$. \cite{Paul,Paul1,Grest}
A subdiffusive regime has been observed also
in melts of entangled
polymer chains for time shorter than the entanglement time, $t_e$ (for $t \geq t_e$ 
the polymer motion
is restricted to a curvilinear diffusion) where the conditions for Rouse dynamics -
purely diffusive behavior - would seem to hold. 
In polymer solutions, the subdiffusive dynamics  
only appears
above the overlap concentration, $c^*$\cite{Kremer,Hall},
suggesting that this effect is driven by intermolecular forces.\cite{Paul,Paul1}

The failure of the Rouse approach in the short-time regime is actually not surprising.
The underlying assumption of this model is that the dynamics 
of the surrounding fluid, and hence
the intermolecular 
time correlation functions, relax on a time
scale much shorter than any  of the
intramolecular processes. This allows the minimization of the memory contribution
resulting from the projection of the fluid 
dynamics onto the
intramolecular coordinates, defined as the slow variables in 
the system.\cite{Hansen}
In polymer melts and concentrated polymer solutions, this assumption
obviously does not hold, since there is no
separation of time scales between intramolecular and intermolecular relaxations.
The characteristic time of intermolecular relaxation,
$\tau_{inter}$, is defined as the time required for the polymer
to escape from the influence of the 
intermolecular potential of mean-force.
In polymer fluids the range of the potential
is given by the extension of the
"correlation hole"\cite{DeGennes} in the pair-distribution function, which is
of the order of the molecular radius-of-gyration,
$R_g$.
Thus, 
$\tau_{inter}$ appears to be of the same order of magnitude of the longest
intramolecular relaxation time, $\tau_{inter}\approx\tau_{Rouse}$.
As a consequence in 
the time region $t \leq \tau_{Rouse}$, where the intermolecular correlation functions 
have not yet decayed to zero,
we expect the dynamics to be substantially
affected by the intermolecular effects, and the Rouse equation
should not hold.

The correction due to intermolecular effects becomes more relevant in melts of 
high molecular weight polymer chains since it depends on
the number of interacting polymers to be found on average within
the range of the potential, $n \approx \rho l^3 \sqrt{N}$.
Intermolecular effects also increase with 
the number of molecules undergoing cooperative
dynamics, when the system approaches its glass transition.\cite{glasses}

While the theoretical approach
to the structure of the macromolecular fluid has been successfully implemented
through a rigorous formalism
to include at the same level of detail both intramolecular and intermolecular
interactions\cite{prism}, the rigorous, non phenomenologycal, upgrade of the dynamics 
appears quite challenging.
Our approach is in this direction.\cite{CDGLE} We derive a GLE from the 
first-principles Liouville equation
by projecting the dynamics\cite{hansen} 
on a set of variables given by the coordinates
of the molecules undergoing slow
cooperative dynamics.\cite{CDGLE} 
Polymer melts are dynamically heterogeneous, and it is possible to observe 
interconverting regions of slow and fast dynamics.\cite{Ediger}
When projecting on the group of slow interacting polymers, the projection approximation,
together with the mean-field description of the liquid of fast molecules, becomes
well justified. 
The derived GLE describes the dynamics of a polymer chain inside a dynamically 
heterogeneous melt.
The equation
explicitly includes intramolecular and intermolecular forces 
in both the frequency term and the memory function.
The traditional single-polymer GLE is recovered when 
the fluid is described as a mean-field continuum, i.e. the pair distribution
function $g(r)=1$.
In the opposite limit of two interacting
structureless particles, the equation 
recovers the well known GLE, extensively studied in the
past.\cite{2particles} Thus, the equation holds in both the well-known opposite limits of 
dominant intramolecular or intermolecular interactions.

Starting from the Cooperative Dynamics Generalized Langevin Equation (CDGLE), 
we investigate here the anomalous c.o.m. dynamics for an
unentangled polymer chain inside a dynamically heterogeneous 
melt. In section II we present
an  analytical Gaussian core expression for 
the intermolecular c.o.m. potential between two interacting polymers which
allows an analytical solution of the GLE through a normal modes analysis.
Anomalous scaling
exponents and asymptotic values for the c.o.m. dynamics are 
presented in Section III, together with
numerical model calculations. The comparison of
the theory with 
computer simulation data of unentangled  c.o.m. dynamics follows
in Section IV.
A brief discussion concludes the paper.

\section{Center-of-mass cooperative dynamics, and intermolecular potential}
We start from the CDGLE for the dynamics of a single polymer chain in a 
structurally homogeneous, but dynamically heterogeneous fluid. An effective 
segment of index $a$,
in a polymer of index $i$,
follows\cite{CDGLE} 
\begin{eqnarray}
\zeta_{0} \frac{d{\bf r}_a^{(i)}(t)}{dt} & = &
\frac{1}{\beta} \frac{\partial}
{\partial {\bf r}_a^{(i)}(t)}
\ln [\prod_{j=1}^n \Psi({\bf r}^{(j)}(t)) \nonumber \\
&& \prod_{k<j}^n g({\bf r}^{(j)}(t)-{\bf
r}^{(k)}(t))] -\frac{\beta}{3}\sum_{b\neq a}^{N} \int_0^t d \tau
<{\bf F}_a^{(i)} \cdot {\bf F}_b^{Q(i)}(t-\tau)>
\frac{d{\bf r}_a^{(i)}(\tau)}{d\tau} \label{CDGLE} \\
&& - \frac{\beta}{3} \sum_{b=1}^{N}\sum_{j\neq i}^n \int_0^t d \tau
<{\bf F}_a^{(i)} \cdot {\bf F}_b^{Q(j)}(t-\tau)> \frac{d{\bf r}_a^{(i)}(\tau)}{d\tau}
+ {\bf F}_a^{Q(i)}(t) \ \nonumber ,
\end{eqnarray}
where
$\Psi({\bf r}^i(t))$ is the intramolecular distribution function,
$-\beta^{-1} ln g[{\bf r}^j(t),{\bf r}^k(t)]$ is the
intermolecular potential of mean-force between two sites (monomers)
belonging to two different polymers, and $\beta=(k_B T)^{-1}$.
${\bf F}_a^{Q(i)}(t)$ is the projected random force ($Q=1-P$, where $P$
is the projection operator\cite{hansen,CDGLE}) that represents the
random collision with the surrounding fluid.
This Generalized Langevin equation is non linear, and couples the 
dynamics of the interacting
chains through the intermolecular contributions included in the frequency
term and in the memory function. The equation cannot be solved analytically, unless a
set of approximations is introduced.

For unentangled polymer melts the equation can be simplified, since  
the corrections due to the memory function contributions can 
be in first approximation
neglected for several reasons.
Because our approach derives the Langevin equation by projecting the Liouvillian 
dynamics over the extended
set of basis functions (not only {\it intram}olecular polymer coordinates as in the 
derivation of the Rouse equation, 
but also {\it inter}molecular coordinates)
the memory function corrections are further minimized with 
respect to the conventional
intramolecular single chain description.\cite{hansen}
The purely intramolecular
equation reduces to the Rouse equation when memory function contributions 
are neglected.\cite{BixonZw} 
The Rouse
equation is already a reasonable approximation to the single polymer dynamics,
and is correct in the long-time regime. It is commonly accepted that when
the memory function contribution is included, the intramolecular diffusive
equation describes 
the cross-over to entangled
polymer dynamics.\cite{pmc,Hess}

When the memory function contributions are neglected and we focus on the
single chain center-of-mass dynamics for an unentangled polymer chain
of index $i$,
Eq.(~\ref{CDGLE}) reduces to
\begin{eqnarray}
\zeta \frac{d{\bf r}_{c.m.}^{(i)}(t)}{dt} & = &  
\frac{1}{\beta} \frac{\partial}
{\partial {\bf r}_{c.m.}^{(i)}(t)}
\ln [\prod_{j=1}^n 
\prod_{k<j}^n g({\bf r}_{c.m.}^{(j)}(t)-{\bf
r}_{c.m.}^{(k)}(t))] 
+ {\bf F}_{c.m.}^{(i)}(t) \  \label{pill},
\end{eqnarray}
since the c.o.m. diffusion is decoupled from the internal relaxation dynamics.
$-\beta^{-1} ln g[{\bf r}^j(t),{\bf r}^k(t)]$ is the 
intermolecular potential of mean-force between the c.o.m.
belonging to two different polymers, 
$\zeta$ is an effective friction coefficient, and
$<{\bf F}_{c.m.}^{(i)}(t)\cdot {\bf F}_{c.m.}^{(j)}(0)>=6 
\zeta \beta^{-1} \delta(t)$ 
the fluctuation-dissipation condition.

\subsection{Effective Gaussian-core potential between two interacting polymers}
To pursue on analytical solution of Eq.(~\ref{pill}) we derive an analytical, Gaussian-core 
expression for the
intermolecular potential 
between the c.o.m. of a pair of polymers.
The modeling of
polymer coils
as interpenetrable soft spheres with a realistic Gaussian-core potential
has been a subject of long-standing interest. In addition to the obvious scientific
relevance\cite{Stillinger}, such an approach  could
significantly reduce the computational time required in the simulations
of polymer fluids.\cite{Kremer1} 
For polymer solutions, a mean-field derivation
of the potential for
interpenetrating polymers
was obtained 
by Flory and Krigbaum.\cite{Flory} Their mean-field approach predicts a 
potential at contact that increases with
$N$, in disagreement with computer simulations where 
the contact potential decreases with increasing $N$.
An alternative derivation through scaling theory\cite{scaling} and renormalization
group calculations\cite{RG}, recovers the correct scaling exponent in the asymptotic
$N \rightarrow \infty$ regime. In a recent series of papers Hansen and coworkers
presented an
empirical form of the potential in the finite-size regime\cite{Hansenp},
and showed, by comparison with computer
simulations, that the Gaussian-core model gives a very 
good description of the
interpolymer potential in a wide range of densities and temperatures. 
The same study also pointed out that interactions of higher order
than the effective pair interactions do not give relevant contributions to
the calculation of the thermodynamic properties of the fluid
in the semidilute regime.

We derive here a Gaussian-core pair potential 
for the interaction between the c.o.m. of two interacting polymers in
a melt. 
The potential 
is calculated as a function of the c.o.m. interpolymer
distance, $R(t)=|{\bf R}(t)|$. If ${\bf r}_a^{(i)}$ is the position of monomer 
$a$ in chain $i$ with respect
to its c.o.m. coordinates, and
${\bf r}_b^{(j)}$ is the position of monomer $b$ in chain $j$ in the same coordinate system, 
the distance between $a$ and $b$ is given by ${\bf r}={\bf r}_b^{(j)}
+ {\bf R}(t) - {\bf r}_a^{(i)}$. 
At any instant the segments are statistically
positioned around the c.o.m. following a Gaussian distribution\cite{Flory1}
\begin{equation}
\Psi(r_{a})=\left ( \frac{3}{2 \pi R_g^2} \right )^{3/2} e^{\frac{-3
{\bf r}_{a}^2}
{2 R_g^2}} \ .
\end{equation}
The distance between the polymers' c.o.m.  is exactly obtained from the integral
${\bf R}= \int d {\bf r}_a^{(i)} \Psi(r_a^{(i)})
\int d {\bf r}_b^{(j)} \Psi(r_b^{(j)}) {\bf r}$.

In analogy with the previous calculation, the c.o.m. pair distribution function, 
$g(R)$, is calculated by 
averaging over the monomer positions, the four-point distribution function
which correlates two monomers on different polymers, and their c.o.m..
The four point distribution function is approximated by the
product of three two-point distribution functions according to
$g(R) \approx \int d {\bf r}_a^{(i)} \Psi(r_a^{(i)})
\int d {\bf r}_b^{(j)} \Psi(r_b^{(j)}) g(r)$.
For the monomer pair distribution 
function,  we adopt the
polymer thread-model
form, which has been shown to reproduce many fundamental
features of the polymer melt structure\cite{prism}, 
\begin{eqnarray}
g(r)=1+ \xi_{\rho}\left[\frac{e^{-r/\xi_{\rho}}}
{r} - \frac{e^{-r/\xi_c}}{r}\right] \ .
\end{eqnarray}
Two characteristic lengths enter the equation: the local
density screening length $\xi_{\rho}=3l/(\pi \rho^*)$ with $\rho^*=\rho l^3$ the reduced
fluid density, and
the correlation hole length scale  $\xi_c=R_g/\sqrt{2}$.
In the "thread" model 
the  chain is composed of beads of
vanishing thickness, while the "bare" interpolymer interactions are
simple
hard core
repulsions accounted for by the condition that the site-site pair correlation
function vanishes at contact [$g(d)=0$], while the intermolecular direct
correlation function
$c(r)=c_0 \delta(r)$.

We solve the integral introducing the Fourier transform
\begin{eqnarray}
\frac{1}{r}=\frac{4 \pi}{(2 \pi)^3} \int d {\bf k} \frac{e^{i {\bf k}\cdot {\bf r}(t)}}{k^2} \ ,
\end{eqnarray}
and integrating over the coordinates of the two monomers.
The c.o.m. pair distribution function 
\begin{eqnarray}
g(R) \approx 1 + \xi_{\rho}\frac{2}{\pi} \frac{1}{R}
\int_0^\infty d k k  \sin(kR) \left[ \frac{\xi_{c}^{-2}-\xi_{\rho}^{-2}}
{(\xi_{\rho}^{-2} + k^2)
(\xi_{c}^{-2} + k^2)} \right] e^{-k^2 R_g^2 /3}
\ .
\end{eqnarray}
The exponential terms are dominat for $k^2 \leq N^{-1}$, which implies 
$k^2 \xi_{\rho}^2 \leq N^{-1}$ and $k^2 \xi_c^2 \leq 1$. This approximation
simplifies the previous equation to
\begin{eqnarray}
g(R) \approx 1 + \xi_{\rho} \left[\xi_{\rho}^2 - \xi_c^2\right] \sqrt{\frac{2}{\pi}}
\left(\frac{3}{2 R_g^2} \right)^{3/2} e^{-\frac{3R^2}{4 R_g^2}}
\end{eqnarray}
By keeping the first order term 
in the expansion of the logarithm and introducing the definitions of 
$\xi_{\rho}$ and $\xi_{c}$,
the approximate solution for the interpolymer potential between two polymers
of large, finite $N$ becomes
\begin{eqnarray}
w(R) = - \log g(R) \approx \frac{27 \sqrt{2}}{4 \pi \sqrt{\pi}} \frac{1}
{\sqrt{N}\rho^*} 
[1 - 108 \pi^{-2} (\rho^*)^{-2}(N)^{-1}] e^{-3 R^2/(4 R_g^2)} \ . \label{pot}
\end{eqnarray}
$w(R)$ is Gaussian and finite at all interpolymer
distances, with a range of the order of the radius of gyration.
At full polymer-polymer
overlap,
$w(0)$ decreases with increasing reduced fluid density,
$\rho^*$, reflecting the transition from compressible to incompressible
systems. It also decreases with increasing
polymer molecular weight due to the increasing interpenetrability of the polymer
chains. 
Eq.(~\ref{pot}) is in qualitative agreement with the trend observed
in computer simulations of semidilute polymer solutions.
The last section of the paper shows that Eq.(~\ref{pot}) is in good agreement
with data from
computer trajectories provided by Grest\cite{Grest} for polymer melts at different
temperature, density, and molecular weight.

The derived interpolymer potential belongs to the class of bounded interaction potentials 
that do not diverge at the origin.\cite{Stillinger} Such potentials arise naturally as 
effective interactions between the c.o.m. of soft, flexible macromolecules
since the c.o.m. of two molecules can coincide without violation of the excluded 
volume conditions. 
At complete polymer overlap, the $N \rightarrow \infty$ scaling limit of $w(0)$ 
is not finite
in disagreement with the observed behavior in semidilute solutions. However a comparison
with computer simulations for the melt is not possible in this study since our data
do not reach the $N \rightarrow \infty$ scaling regime.

\subsection{Many-chain center-of-mass dynamics}
To include the  pair intermolecular potential just derived in Eq.(~\ref{CDGLE}),
several steps need to be undertaken. 
Since it is the distinct part of the Van Hove function that appears in 
Eq.(~\ref{CDGLE)}, we assume a 
time-dependence of the potential, which enters
through the evolution of the square intermolecular distance between the 
c.o.m. of the two polymers, $R^2(t)$.
The intermolecular term in Eq.(~\ref{CDGLE}) results from the decomposition of the
$n \ N$ body distribution function in the product of intra- and inter-molecular
pair distribution functions.\cite{CDGLE} In principle this high order distribution function
is not equivalent to the product of two-body terms. However a decomposition in
pair potentials is the natural approximation to adopt,
and in our case it is  supported by the quantitative agreement
with the simulation data.
The product of the pair distribution functions gives $n$ equivalent terms
in Eq.(~\ref{pill}). The intermolecular force for a group
of $n \approx \rho^* \sqrt{N}$ interacting polymer chains is
\begin{eqnarray}
G(t) R(t)
\approx \frac{81 \sqrt{2}}{8 \pi \sqrt{\pi}} R_g^{-2}
[1 - 108 \pi^{-2} (\rho^*)^{-2}(N)^{-1}]
e^{-3 <R^2(t)>/(4 R_g^2)} R(t)
 \label{gt} \ .
\end{eqnarray}
The intermolecular distance is
approximated by its statistical average over the polymer configurational space,
$<R^2(t)>$, and in the many-chain description becomes
$<R^2(t)>\approx n \Delta R^2(t)- 6 D_{\chi} t$, where $\Delta R^2(t)$ is the single
chain mean-square displacement. $D_{\chi}=D_{Rouse}/(\rho^* \sqrt{N})$ 
is the collective
diffusion coefficient derived below.
The approximate effective potential that corresponds to the force in Eq.(~\ref{gt}),
is found to be in good agreement with the time-dependent
potential from computer simulations, as discussed in the last section of the paper.

Eq.(~\ref{pill}), with Eq.(~\ref{gt}), reduces to a set of coupled equations
of motion for the c.o.m. polymer dynamics
\begin{eqnarray}
\zeta \frac{d {\bf r}_{c.m.}^{(i)}(t)}{d t} = 
G(t) [ {\bf r}_{c.m.}^{(i)}(t)  - R^{cm}(t)]
+ {\bf F}_a^{Q(i)}(t)  \ ,
\label{rai}
\end{eqnarray}
where $R^{cm}(t)=n^{-1} \sum_{j=1}^n {\bf r}_{cm}^{(j)}(t)$ is the coordinate
of the c.o.m. position for the dynamical heterogeneity.
Eq.(~\ref{rai})  has the
advantage of including both the intramolecular
and the intermolecular forces  while
keeping the mathematical simplicity of the Rouse equation, and
it can be solved analytically by 
transformation into normal
modes of motion. 

Because of the symmetry of the system, composed by equivalent
polymer chains,
the set of coupled equations of motion reduces to two 
independent equations in the
relative,
${\bf r}_\xi(t)=2^{-1/2} [{\bf r}_{cm}^{(i)}(t)-{\bf r}_{cm}^{(j)}(t)]$,
and collective 
coordinates, 
${\bf r}_\chi(t)=n^{-1/2} \sum_{i=1}^n {\bf r}_{cm}^{(i)}(t)$.
\begin{eqnarray}
\zeta \frac{d {\bf r}_\xi(t)}{d t}& = &  G(t)
{\bf r}_\xi(t)
+ {\bf F}^{\xi}(t) \ , \label{xieq}\\
\zeta \frac{d {\bf r}_\chi (t)}{d t}& = & 
{\bf F}^{\chi}(t) \ . \label{chi}
\end{eqnarray}
The relative and collective projected force correlation functions in the Markov
limit obey the fluctuation-dissipation conditions
\begin{eqnarray}
<{\bf F}^{\xi}_\alpha (t) \cdot {\bf F}^{\xi }_\gamma(t')> = 6
\zeta \beta^{-1} \delta(t-t') \delta_{\alpha \gamma} \ , \ \ \ 
<{\bf F}^{\chi }_\alpha (t) {\bf F}^{\chi }_\gamma(t')> = 6
\zeta \beta^{-1} \delta(t-t') \delta_{\alpha \gamma} \ .
\end{eqnarray}
Eqs.(~\ref{xieq},~\ref{chi}) are exactly solved through standard 
procedures\cite{Mantegna,Fokker},
and the 
single chain c.o.m. coordinate,
in the ensemble of molecules undergoing
correlated
dynamics, is obtained from the combination of the collective and relative contributions.

\section{Center-of-mass dynamics:
asymptotic behavior and model calculations}
The single-chain mean-square displacement of the c.o.m.  
becomes
\begin{eqnarray}
\Delta R^2(t)& = & < \left( {\bf R}_{c.m.}(t)-{\bf R}_{c.m.}(0) \right)^2> = \frac{n-1}{n} 
< \left({\bf r}_\xi(t)- {\bf r}_\xi(0)\right)^2 > + \frac{1}{n}
< \left({\bf r}_\chi(t)-{\bf r}_\chi(0) \right)^2 > \ , \label{cmmsd} \\
& \approx & < \left({\bf r}_\xi(t)- {\bf r}_\xi(0)\right)^2 > + 6 D_{Rouse} t /
(\rho \sqrt{N})
\ . \nonumber
\end{eqnarray}
While the collective contribution is linear in
time with a constant diffusion coefficient $D_{\chi}$,
the relative contribution exhibits a more
complex behavior that depends on the strength of the intermolecular 
interaction, $G(t)$. From Eq.(~\ref{gt}) $G(0)$ is defined as the force at complete
polymer overlap.
\begin{eqnarray}
<({\bf r}_\xi(t) - {\bf r}_\xi(0))^2 > & = & 
<{\bf r}_\xi^2(0)> \left( e^{ - \alpha(t)}
-1 \right)^2 + 6 D_{Rouse}
e^{ - 2 \alpha(t)}
\int_0^t e^{2 \alpha(t')} dt' \ , \nonumber
\end{eqnarray}
where
\begin{eqnarray}
\alpha(t)=\frac{ G(0)}{\zeta} \int_0^t e^{-\frac{3 <R^2(t')>}
{4R_g^2}} dt' = \zeta^{-1} \int_0^t G(t') dt' \ . \label{alpha}
\end{eqnarray}
Interpolymer interactions vanish at long time, $t \gg \tau_d$, where we define
$\tau_{d}$ as the characteristic time in which $G(t) \rightarrow 0$.
In general
$\tau_d$ describes the time that a molecule has to travel to escape from the
range of the intermolecular potential.
In uncharged polymer fluids at temperature well above their glass transition,
the range of the potential is comparable to $R_g$, and $\tau_d \approx \tau_{Rouse}$.
For $t \gg \tau_{d}$, the single-chain c.o.m. mean-square displacement 
recovers Rouse diffusion.
For $t \ll \tau_{d}$, 
$G(t)$ is a function of intermolecular distance $<R^2(t)>$
and Eq.(~\ref{cmmsd}) has to be solved numerically
through a self-consistent procedure. 

In a few well-defined conditions Eq.(~\ref{cmmsd}) is analytically
soluble.
At short time the single-chain c.o.m. mean-square displacement
follows 
$\Delta R^2 (t) \approx
6  t/(\beta N \zeta_{0})+ G(0)^2 <R^2(0)> t^2/(\zeta_{0}^2)$. 
If $G(0) \rightarrow 0$ it recovers the conventional Langevin dynamics,
while, if $|G(0)| > 0$, it recovers a short-time superdiffusive behavior
characteristic of Langevin equations with a time dependent
drift coefficient.\cite{Mantegna,Fokker} Since Langevin equations only describe
overdamped chain dynamics, they do not hold in the short-time region where
the dynamics is precollisional.\cite{Balucani} 
In the present study, we focus on the 
subdiffusive and free-diffusive c.o.m.
dynamics in the postcollisional regime.

The onset of the 
anomalous subdiffusive regime takes place  
at the characteristic time $\tau_1 = \frac{\zeta}{2 G(0)}$. 
This suggests that a system
with intermolecular
interaction  of increasing strength 
"freezes"  its
dynamics at shorter and shorter
times.
Equivalently we can say that the motion
begins to be suppressed on the
largest length scale (smallest index modes of motion), and by 
increasing $G(0)$ the dynamics is progressively frozen at
shorter and shorter length scales, or higher and higher 
index modes.\cite{modes}
The increase of the monomer relative friction coefficient has an opposite 
effect, since it slows down the dynamics
and enforces the overdamped regime.

For $t \gg \tau_1$ the
dynamics are different depending on the nature of the long-range potential.
Since the c.o.m. interpolymer potential results from the combination of
monomer-monomer interactions, and in polymer melts excluded volume interactions
are dominant, this potential is repulsive.
An
effective intermolecular
long-range attractive component can appear, however, when the single-polymer is 
constrained,
for example because of the
presence of chain-chain uncrossability or entanglements. In those cases, the molecule
has to overcome an effective potential barrier in order to diffuse. Attractive effective
potential are also presents in polymer blends, self-assembling polymer systems, and in
droplets of polymer
liquids.

If we approximate the infinitesimal mean-square-displacement as linear in time,
the interpolymer distance
$<\Delta R^2(t)> \approx 6 D t$. When
introduced in Eq.(~\ref{alpha}) it yields
\begin{eqnarray}
\alpha(t)=\frac{2 G(0) R_g^2}{9 \zeta D }
\left(1-e^{-\frac{9 D t}{2 R_g^2}} \right) \ .
\end{eqnarray}
For short, highly mobile polymer melts, $9 D t \gg 2 R_g^2 $,
the dynamics is simply diffusive at any time-scale.

In a melt of slowly moving, long polymer chains, $9 D t \ll 2 R_g^2 $.
If the potential is repulsive, 
the single-chain dynamics 
crosses over from the 
subdiffusive regime, which
describes the slowing down of the single-chain dynamics due to the
presence of the surrounding molecules ("cage effect"), to free single-chain
diffusion
at $t \approx 2 R_g^2 /( 9 D) $, where
$\alpha(t)$ becomes constant, $\alpha(t) \approx - 2 G(0) R_g^2/(9 D \zeta)$.
A decrease in the polymer molecular weight, or an increase in the chain mobility
decreases the duration of the subdiffusive regime.

The c.o.m. dynamics of slowly moving, long polymer chains
interacting through an attractive
long-range potential, follows a subdiffusive regime at intermediate time.
In strongly interacting systems the relative mean-square displacement is constant,
$<({\bf r}_\xi(t) - {\bf r}_\xi(0))^2 >  = 
<{\bf r}_\xi^2(0)> +
 3/[N\beta G(0)]$, and
the single-chain dynamics is driven by the collective contribution.
The surrounding molecules act as a cage of the tagged
molecule dynamics.
At longer times, when $t \approx 2 R_g^2 /(9 D)$, 
the single-chain exhibits free
Fickian dynamics. This regime corresponds to the molecule's excape from
the surrounding "cage", as it overcomes 
an effective energetic potential barrier.  If the time scale of observation is
shorter than this dissociation time,  the c.o.m. interdiffusion 
is frozen, and the system behaves non-ergodically.

The asymptotic behaviors just discussed are illustrated  by numerical 
calculations in Figures 1 and 2.
We investigate the influence of the intermolecular force, $G(t)$, in
the scenarios of attractive and repulsive long-range interactions for
long,  slowly-moving polymer fluids,
$9 D t \gg 2 R_g^2 $. In short highly mobile polymer chains the anomalous
dynamics is less relevant.

Eq.(~\ref{rai}), when the definition of $G(t)$ in Eq.(~\ref{gt}) is introduced, 
becomes non linear as the 
intermolecular potential 
depends on the interpolymer distance, $R^2(t)$, that changes 
while the system evolves in time.
Thus the short-time dynamics depends on the choice of the
initial interpolymer distance.
Different values of the distance, however, mainly affect the dynamics in the
precollisional regime that is
outside the range of validity of our equation.
An estimate of the initial interpolymer distance, which is 
a function of the bulk properties of the fluid, can be obtained from
the distribution in space of the polymer chains inside the volume spanned by the
potential. The average number of chains in a sphere of radius $r$, 
as a function of time, is given by the time-dependent pair distribution
\begin{eqnarray}
n[r(t)]= N^{-1}\int_0^{r} d{\bf r'} \rho g[r'(t)] \ ,  \label{n}
\end{eqnarray}
which gives for the initial mean-square intermolecular distance
inside the range of the potential
\begin{eqnarray}
<R^2 (0)> \propto \int_0^{\sim R_g} n[r(0)]r^2 dr \approx R_g^2 \ . \label{R}
\end{eqnarray}
Since the
intramolecular monomer distribution is Gaussian
the chains inside the potential range are not 
uniformly distributed,
and the resulting average intermolecular distance is of the order 
of the polymer radius-of-gyration.

To perform quantitative model calculations we specify 
the numerical value of a few more parameters. 
The model polymer chain 
comprises $N=100$ statistical segments
of unitary length $l$, with the chain described by a 
freely jointed model consistently with
Rouse theory.
The effective temperature is set to $k_B T=1$ in units of $G(0)$, and we assume the
density of the melt state, $\rho =1$.
The number of chains initially correlated inside the range of the potential
is given by Eq.(~\ref{n}) as
$n\approx 4/3 \pi \sqrt{N} l^3 \rho$. $n$ is also the number of chains 
undergoing correlated
motion, defined from Eq.(~\ref{CDGLE}) as the number of chains interacting
through the mean-force potential.
The effective time-independent monomer
friction coefficient 
is set to $\zeta=1$.

In Figures 1 and 2, we
study the decay in time of the interaction strength, $G(t)$, when its 
value at complete polymer overlap, $G(0)$, is varied for a system of slowly moving, 
high molecular weight polymers. We use as an effective parameter the 
interaction strength, $G'(0)$, that
is a multiple of $G(0)$ calculated from Eq.(~\ref{gt}).
The features emerging from those plots are related to the onset of anomalous dynamics in 
the c.o.m. mean-square displacement as a function of time. 
In general an increase of the strength of the potential induces the onset of 
anomalous dynamics.

$G(t)$
is normalized by its own value calculated when $<R^2(t)>=R_g^2$, i.e. $G(t)|_{R_g}$.
For small $|G(0)|$, the decay of $G(t)$ follows a single step mechanism that 
corresponds to a mean-square displacement approaching the linear single-chain
diffusion.
For large $|G(0)|$ the behavior depends on the sign of the potential.

For a repulsive intermolecular potential (Figure 1)
an increase of $G(0)$,
for $ t \leq \tau_{Rouse}$, induces the onset of a stretched exponential decay
which crosses over to Rouse behavior at $t \approx \tau_{Rouse}$.
In the c.o.m. mean-square displacement,
the presence of the interactions modifies the Rouse free diffusive behavior by inducing
a subdiffusive behavior at short time.
At 
$t \gg r R_g^2 /(9 D)$  the system recovers free Fickian diffusion.
No many-chain correlated dynamics appears at long time.

To investigate the effect of an attractive effective potential we adopt $G'(0)$ equal
to a multiple of $G(0)$, 
defined by Eq.(~\ref{gt}), assuming a minus sign.
At intermediate time
the system is frozen in a metastable equilibrium
configuration with constant strength of the interactions, and constant
interpolymer distance (see Figure 2).
This regime corresponds to an anomalous subdiffusive
regime ($\Delta R^2(t) \propto t^\nu$ with $\nu \ll 1$). If the thermal energy of the 
system
is high
enough, $\beta G'(0) \rightarrow 0$,
the molecule can escape from the "caged" dynamics
reestablishing the effective ergodicity of the system. Otherwise, the system
undergoes many-chain correlated dynamics, so it
behaves non-ergodically.
Thus, by increasing the strength of the intermolecular
potential
the theory predicts 
an apparent ergodic to non-ergodic transition.

In the Figures are also illustrated the two limiting
cases of a single-chain free dynamics ($G(0)=0$), and for the "caged"
collective dynamics ($G(0)\rightarrow \infty$).
The comparison with those curves emphasizes how the single-chain dynamics in strongly
interacting attractive systems crosses over from the short-time single-chain
dynamics to the collective many-chain diffusion at long time.

\section{Comparison with Simulations}
To test our approach we compare its predictions with 
computer simulations of unentangled polyethylene (PE) dynamics in the melt
state and at decreasing temperature, while approaching the glass transition
(the entanglement degree of 
polymerization $N_e=136$). PE
has been extensively investigated experimentally
\cite{modes,Pearson} because of its industrial applications, and has been 
chosen as a model system for 
computer simulations\cite{Paul,Paul1,Grest}
due to the simplicity of its monomer structure.
We first compare with data from computer simulations 
performed by Grest and coworkers.\cite{Grest}
The simulations calculate the dynamics
at constant volume and constant temperature,
using the experimental
densities at atmospheric pressure, as
reported in Table I.

We use
as an input to the theory the density,
polymer molecular weight, number and length of the statistical segments from
computer simulations (see Table I).
Figure 3 shows very good agreement between theory and simulations, when 
one fitting parameter, i.e. the potential at
complete interpolymer overlap $w'(0)$, is optimized.
Table I reports the ratio between the optimized
analytical values of $w'(0)$ and the theoretical values, $w(0)$, calculated
from Eq.(~\ref{pot}). $w'(0)$ appear to be fairly close to $w(0)$.
The agreement 
with the computer simulations is good for all the
samples considered, which span a wide range of temperatures and
molecular weights.

A second approximation, implicitly introduced in the solution of the
CDGLE, is the averaging of the instantaneous interpolymer distance
in the effective potential, and the self-consistent solution of Eq.(~\ref{rai}).
In Figures 4a, 4b, and 4c we present a comparison between Eq.(~\ref{pot}) and
the 
time-dependent intermolecular
potential between the c.o.m. of two polymers, extracted from computer simulations
as $- \ln g[R(t)]$ for the $C_{30}H_{62}$ sample.
The approximate static interpolymer
potential
appears to reproduce quite well the data at any time scale.
This is because
the time dependence of the potential is quite weak initially, so that
the data 
during the first $30 \ ps$
almost superimpose, and are well reproduced by the same analytical formula.
In Figure 4d we compare Eq.(~\ref{pot}) with the averaged time-dependent Gaussian
potential, corresponding to Eq.(~\ref{gt}), calculated self-consistently with
Eq.(~\ref{rai}). 
The two potentials
show very good agreement.
For time intervals larger than $30 \ ps$, when the potential starts to
differ significatively at short-distance, the molecules have already diffused to large
intermolecular distance and the approximated analytical form of the potential
still reproduces well the data.

From the analytical solution of Eq.(~\ref{rai}) we calculate
the c.o.m. mean-square displacement. The overlap value of the intermolecular potential,
$w'(0)$, is fixed from the previous analysis.
From the long time diffusion
coefficient we derive the value of the effective friction
coefficient, reported in Table I.
Figure 5 shows that the comparison between 
simulations and analytical theory gives an excellent agreement in the whole range
of space and time scales investigated.
This suggests that both the self-consistent procedure
and the approximation of neglecting the memory functions contribution are
reasonable first-order approximations for unentangled melt dynamics.

\subsection{Unentangled undercooled polymer melt center-of-mass dynamics}
We now compare our approach with data from computer simulations of undercooled
polymer systems.
In a recent paper Binder and coworkers\cite{Binder} investigated by computer 
simulations
the onset of short-time anomalous
dynamics for a melt of unentangled
polymer chains.
They performed
Dynamical Monte Carlo computer simulations at decreasing system temperature, 
using the bond fluctuation model\cite{Binderbook} of a chain of $N=10$ statistical segments.
For this model, which is found to reproduce correctly the dynamics of polymer
melts, Binder reports the unit of the
single time step ($1$ Monte Carlo Step $\approx 10^{-13} \ s$) and 
the space unit ($ 1 $ lattice unit $\approx 2.3 \  \stackrel{\circ}{A}$).
\cite{Binderunit}
The chain of $N=10$ effective segments simulated here, corresponds to a polyethylene
chain of
$N\approx 50$ bonds of length $l=1.54 \stackrel{\circ}{A}$.\cite{note1}
Since the model used to simulate the "freezing" of the dynamics privileges
stretched bond configuration, the polymer radius-of-gyration slightly 
increases
with decreasing temperature.\cite{Paulinbinderbook}

We fit
the c.o.m. mean-square displacement having as an input to the equation
the simulation
parameters. 
The monomer friction coefficient is obtained
from the diffusion coefficient in the long-time Rouse regime and is kept
constant
in all the calculations.
We do not include the memory function
corrections\cite{note2} which in our calculations
are negligible for
$N \leq  N_e$.
Because
of the approximations involved in the calculations, we don't strongly rely on the
precision
of the numerical values obtained from the fitting procedure, however  
the theory makes some interesting suggestions on the physics involved in the 
process.

In Figure 6  we show that the fitting procedure gives a reasonable agreement
between the theory and the data
in the complete range of time (length) scale investigated.
At high temperature ($k_BT=1, 0.4$)
the dynamics is driven by repulsive interactions in agreement with our previous
calculations.
The repulsive potential induces a subdiffusive regime
that crosses over to the single chain free diffusion.
In the simulation the decrease of the diffusion coefficient,
follows the empirical Vogel-Fulcher law\cite{Binder}, while the
Rouse equation holds for the monomer local dynamics.
To obtain a good quantitative agreement with the simulations 
we use two fitting parameters: the 
strength of the
potential at complete polymer superposition, $w(0)$, and the number of polymer
chains undergoing cooperative dynamics, $n$.
The best fit with the data is found for $n=13$ and $n=15$, respectively
at $k_BT=1$ and $k_BT=.4$. These values are very close to the theoretically
predicted value.
For temperatures close to the melt state the system behaves
consistently with the study presented above.

When the temperature of the system is further decreased ($k_B T= 0.23, 0.19$)
the subdiffusive behavior appears to have a different
physical origin than in the high temperature limit.
The simulations show an "extremely slow motion", visible also in the
monomer mean-square displacement, which is defined to be a "cage effect".\cite{Binder}
From our
calculations
this regime emerges as a diffusive process involving the correlated dynamics of
many-chains. In this regime the polymer chain is caged
by the presence of the other polymers and cooperative many-chain dynamics
takes place, which indicates an attractive effective potential.
While the monomer friction coefficient is constant from the high temperature
regime, the number of polymer
chains undergoing correlated dynamics increases with decreasing system temperature,
in agreement with experimental studies.\cite{Ediger,Russell,Tracht}
This effect would indicate an increase of the range of the dynamical correlation 
that
would correspond to longer range oscillations in the "real" mean-force
potential and in $g(r)$.
From the fit to the long-time collective diffusion coefficient we obtain 
$n\approx 25$ 
for $k_BT=.23$ and $n\approx 300$
for $k_BT=.19$. These results suggest that the freezing
of the dynamics is related to a rapid increase in the number of polymer
chains involved in the
correlated dynamics.
In this region of temperature
the polymer has to escape from the influence of the intermolecular
potential, overcoming an effective
potential barrier, through a process which is highly cooperative, and this becomes 
increasingly
difficult
at lower
temperature.

\section{Conclusions}
The
presence of effective intermolecular
interactions significantly
modifies the single-chain dynamics 
in the short time, local space regime, $\tau_0 \ll t \ll \tau_{Rouse}$.  
The onset of
anomalous dynamics, observed  in computer simulations of polymer fluids 
in the short-time region where the Rouse equation breaks down, appears to be
quantitatively related to the presence of the effective mean-force potential.
The spatial extent of the anomalous dynamics corresponds 
to the
range of the mean-force potential ($\approx R_g$ in polymer melts well above $T_g$),
while the time scale that characterizes the anomalous dynamics is given by the
time necessary for a polymer to escape from the influence of the intermolecular
potential
($t \approx \tau_{Rouse}$).

The subdiffusive dynamics at time scale $\tau_1\ll t\ll \tau_{Rouse}$,
corresponds to
a slowing down or arrest of the dynamics due to the
presence of intermolecular forces.
It appears to be a cross-over regime
to the long-time diffusive dynamics, consistent with computer simulations
where a wide range of exponents is measured
.\cite{Paul,Paul1,Grest,Kremer,Hall}

The sign of the intermolecular potential qualitatively
modifies the long-range anomalous dynamics.
The theory quantitatively reproduces the experimental data in the complete
range of time and spatial scales investigated.

\section{Acknowledgments}
We are grateful to  G.S.Grest
for sharing the trajectories of his computer simulations.
Acknowledgment is made to the donors of The Petroleum Research Fund,
administrated by the
ACS, for partial support of this research.
We also acknowledge the support of the National Science Foundation under
the grant DMR-9971687.

\section*{TABLE CAPTION}
Table I: Simulation and fitting parameters.

\begin{tabular}{|c|c|c|c|c|c|c|} \hline \hline
$Polymer$ & $T \ [K]$ & $\rho \ [g/cm^3] $ &
$l [\stackrel{\circ}{A}]$ & $w(0)/w'(0)$
& $Initial slope $ & $\zeta \ [10^{9} \ dyn \ s / cm] $ \\
\hline
$C_{10}H_{22}$ & $298$ & $0.7250$ & $3.26$ & $2$ & $0.97$ & $0.25$\\
$C_{16}H_{34}$ & $298$ & $0.7703$ & $3.84$ & $.4$ & $0.94$ & $0.44$\\
$C_{16}H_{34}$ & $323$ & $0.7531$ & $3.75$ & $.6$ & $0.96$ & $0.30$\\
$C_{16}H_{34}$ & $373$ & $0.7187$ & $3.66$ & $.8$ & $0.97$ & $0.18$\\
$C_{30}H_{62}$ & $400$ & $0.7421$ & $4.02$ & $.3$ & $0.90$ & $0.21$ \\
$C_{44}H_{90}$ & $400$ & $0.7570$ & $4.18$ & $.3$ & $0.85$ & $0.27$  \\
\hline \hline
\end{tabular}

\section*{FIGURE CAPTIONS.}

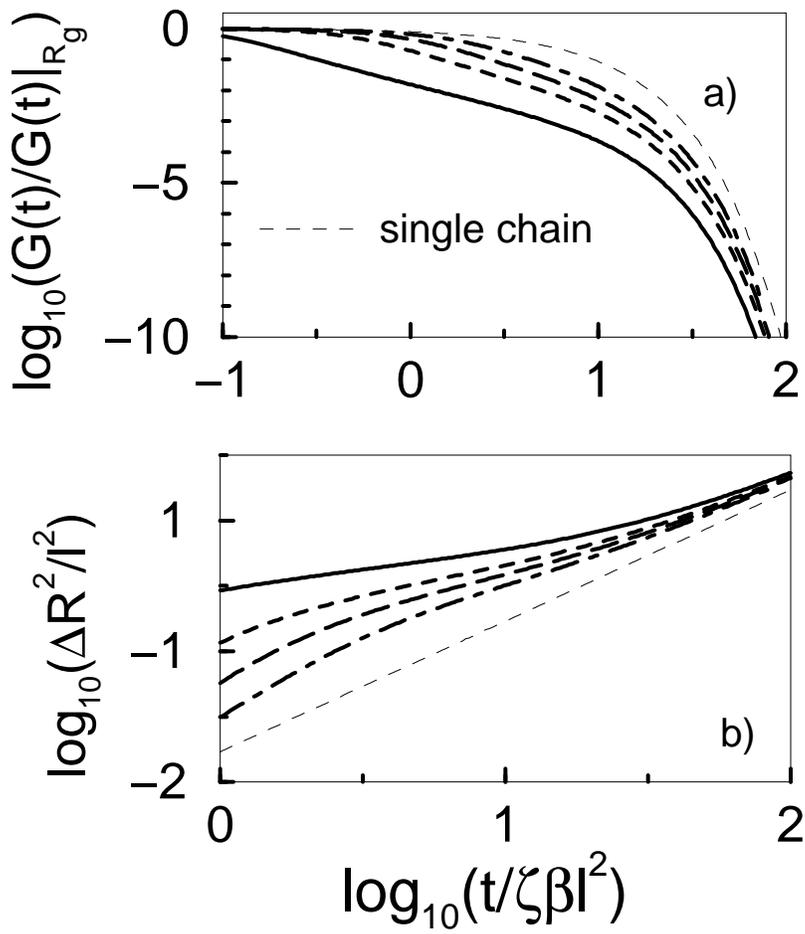
Figure 1: a) Normalized intermolecular constant versus normalized
time, for various  strengths of the repulsive interaction, $G'(0)$.
$G'(0)$ is the effective strength at complete polymer 
overlap from Eq.(~\ref{gt}).
From the top curve (close to the single-chain diffusion) to the
bottom
curve: 
$G'(0)= 0.05 \ G(0)$, $0.1 \ G(0)$, $0.2 \ G(0)$, $G(0)$.
$G(0)$ is the intermolecular interaction at initial conditions:
$k_B T=1$ in units of $G(0)$, $N=100$, $n=\frac{4}{3}\pi \sqrt{N} l^3 \rho$,
$\rho=3 \ l^{-3}$, and $R^2(0)\approx R_g^2$.
Also shown is $G(t)$ for the chain undergoing 
single chain diffusion.
b) Normalized single chain center-of-mass mean-square
displacement vs normalized time, at various strengths of
$G'(0)$, same values that in Figure 1 a).

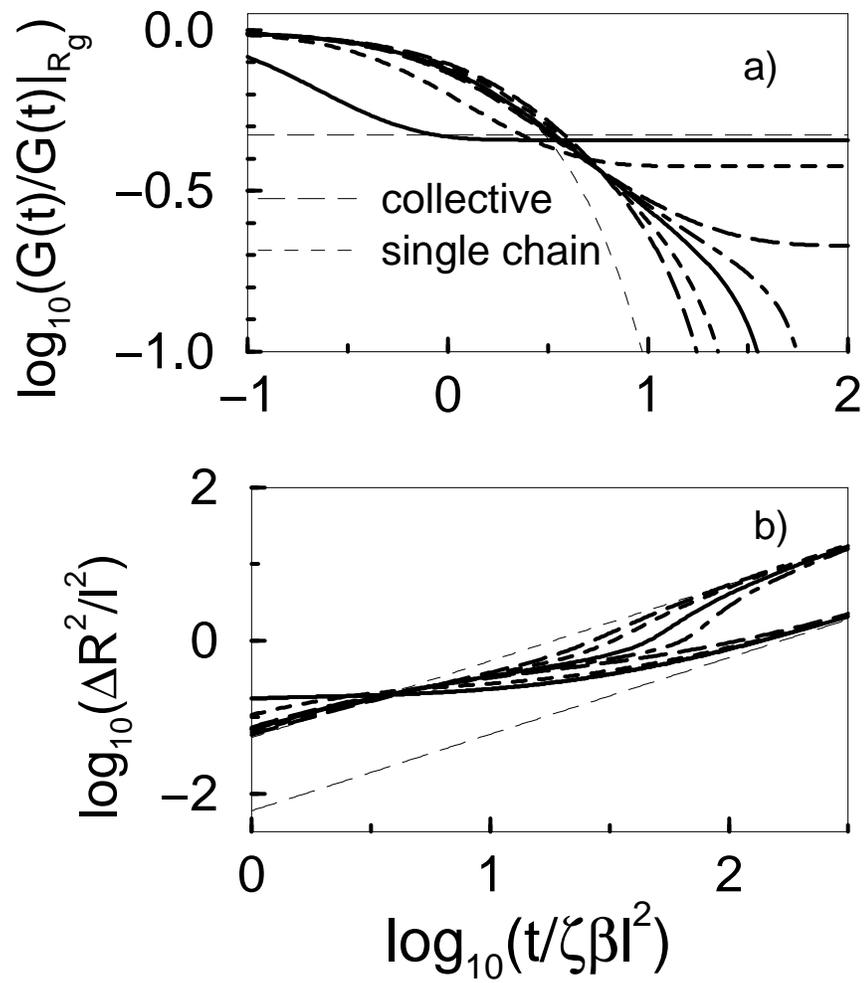
Figure 2: a) Normalized intermolecular constant versus normalized
time, at decreasing strength of the attractive interaction, $G'(0)$.
From the top curve (close to the single-chain diffusion) to the
bottom
curve: $G'(0)=- 0.05 \ G(0)$, $- 0.067 \ G(0)$, $-0.08 \ G(0)$,
$- 0.09 \ G(0)$, $-0.1 \ G(0)$, $-0.2 \  G(0)$, $-G(0)$. 
Same initial conditions than in Figure 1.
Also shown is $G(t)$ for the chain undergoing collective dynamics, and
for the single chain diffusion.
b) Normalized single chain center-of-mass mean-square
displacement vs normalized time, at decreasing $|G'(0)|$, same
values than in Figure 2 a).

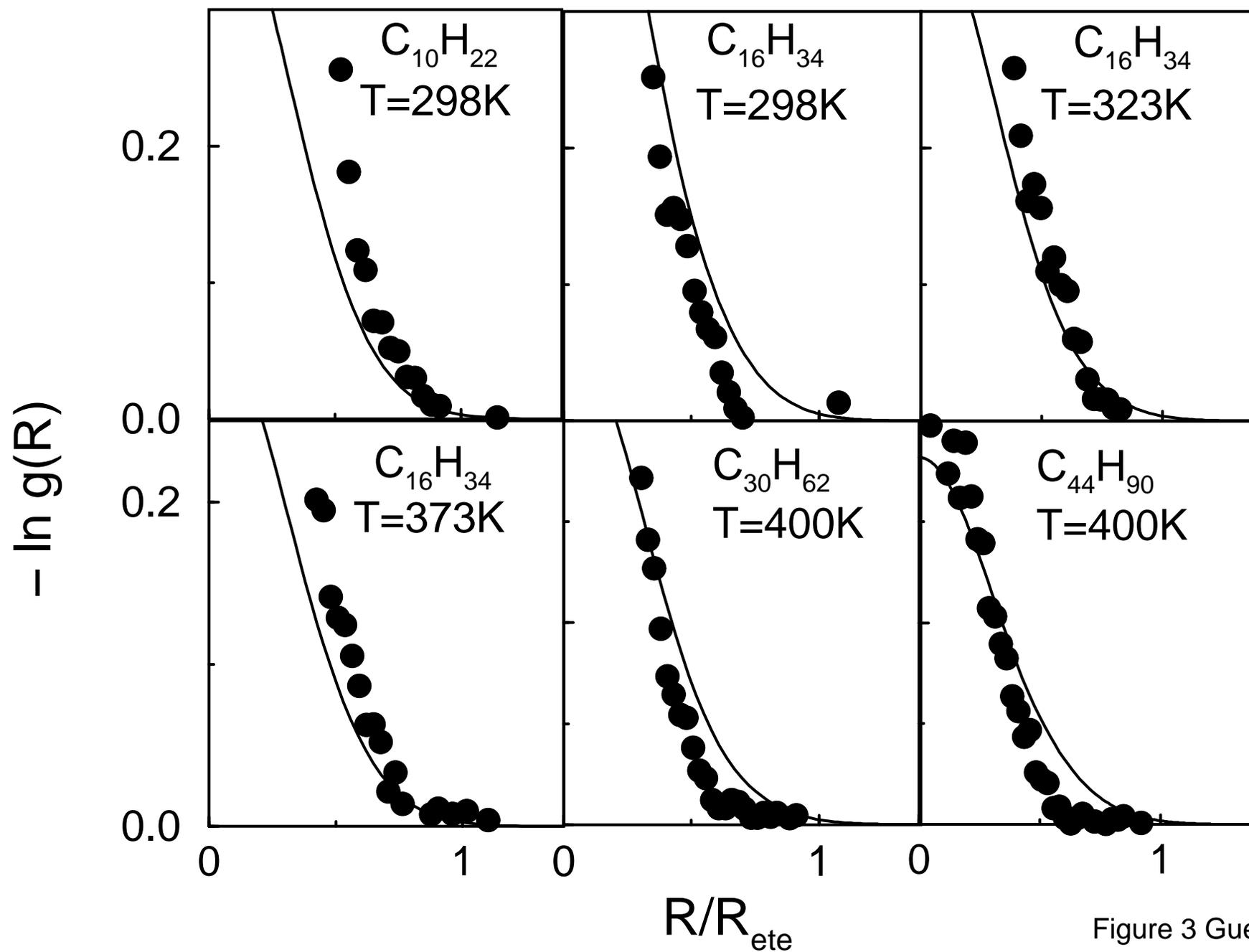
Figure 3: Comparison between the analytical expression of the
soft-core center-of-mass Gaussian potential, Eq.(~\ref{pot}),
and computer simulations data, vs. the c.o.m. interpolymer distance
normalized by the polymer end-to-end distance, $R_{ete}$, from
computer simulations.

Figure 4: Comparison between the analytical equation of the
soft-core center-of-mass Gaussian potential, Eq.(~\ref{pot}),
and the distinct part of the Van Hove distribution function, $- \ln g[R(t)]$, from 
computer simulations of $C_{30}H_{62}$. Plot vs. the c.o.m. interpolymer distance
normalized by the polymer end-to-end distance, $R_{ete}$, from
computer simulations.
Curves at a) $10$ ps, b) $20$ ps, and c) $30$ ps. d) Comparison between Eq.(~\ref{pot})
and the
time-dependent Gaussian potential used in Eq.(\ref{rai}), calculated self-consistently,
for $10$ ps (square), $20$ ps (triangle up), $30$ ps (triangle down).

FIG. 5: Center-of-mass mean-square displacement as a function of time. Best fit of
the molecular dynamics simulation data (filled circle) with the Rouse equation (dashed
line), and with the intermolecular diffusion equation, Eq.(~\ref{rai}), for unentangled
polymer
melts
(full line). The short-dashed lines indicate
the longest Rouse relaxation time, $\tau_{Rouse}$.

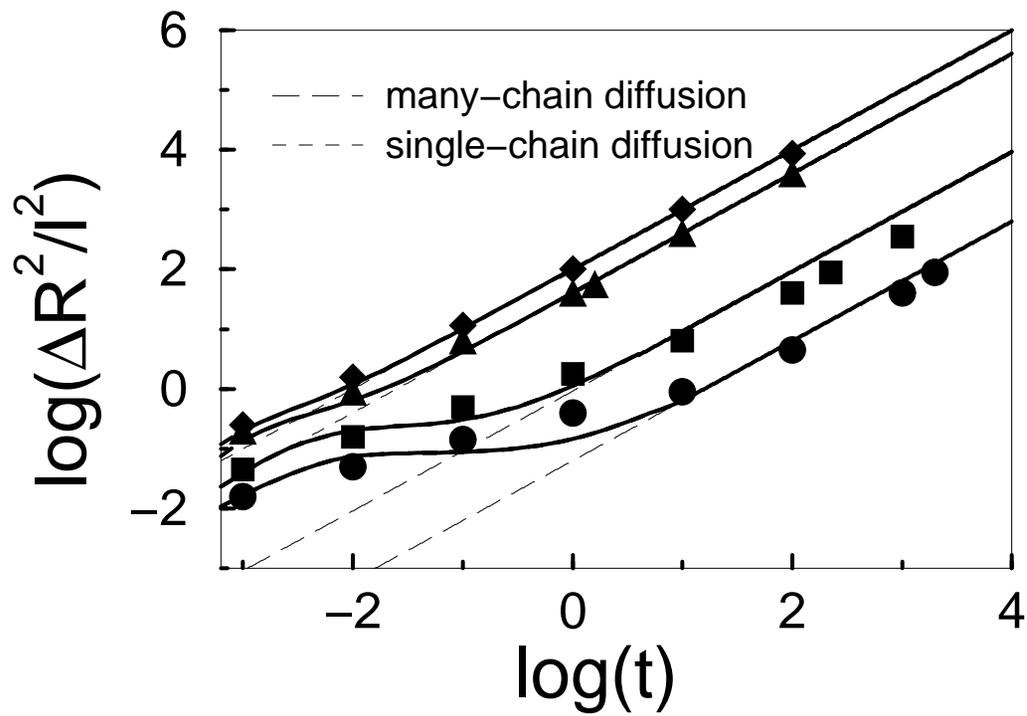
Figure 6: Center-of-mass mean-square displacement as a
function of time at $k_BT=1$ (diamond), $k_BT=0.4$ (triangle), 
$k_BT=0.23$ (square), $k_BT=0.19$ (circle). Best fit of the 
simulation data (ref.\cite{Binder}) with the Rouse equation including 
intermolecular interactions.

\begin{references}
\bibitem{note1} Notice that the bond length in the original figure\cite{Binder}
is an effective segment about five times larger than our $l$.
\end{references}
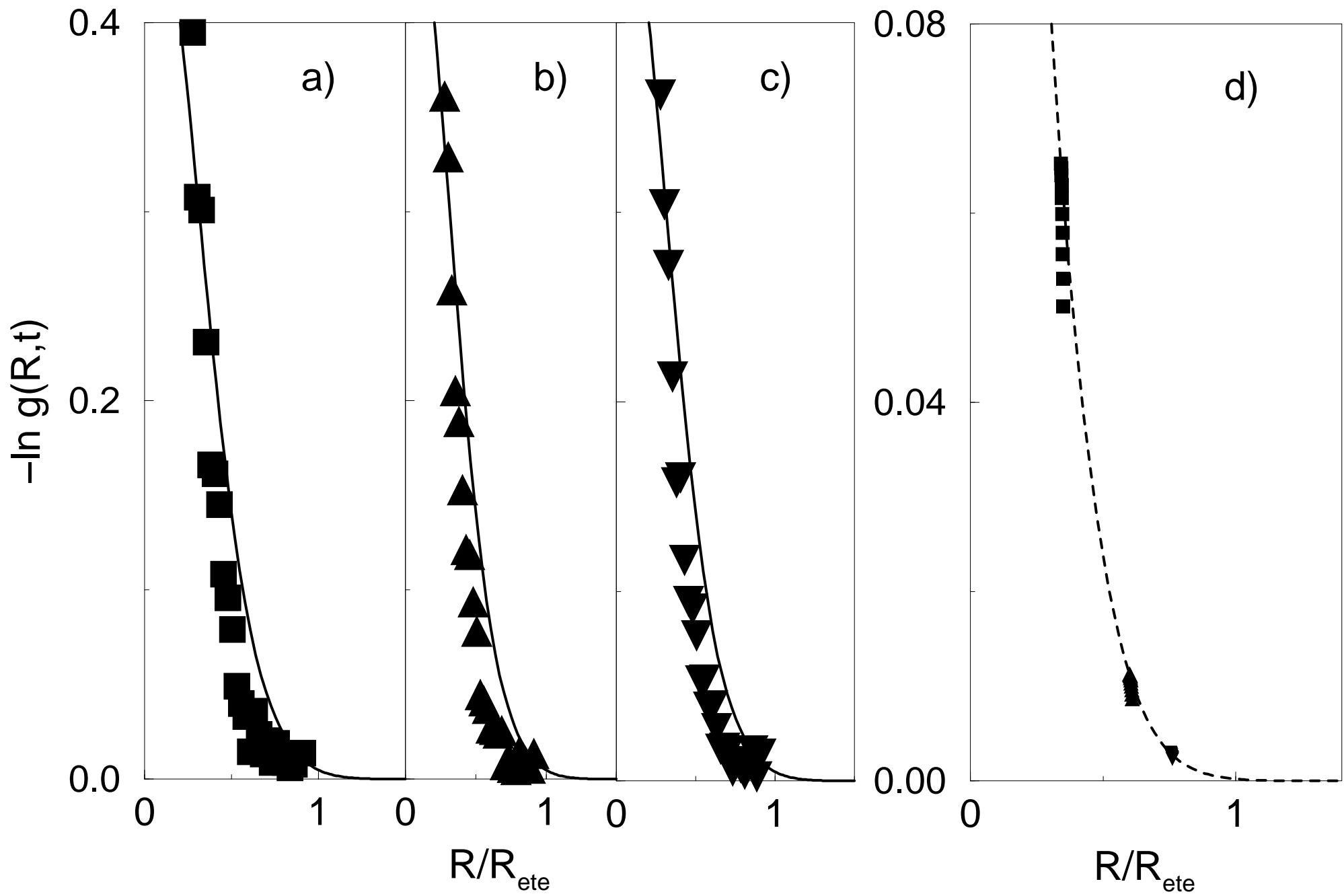
\end{document}